%% file: spl.tex
\documentclass[a4paper,10pt]{article}

\newcommand{\inconly}[1]{}

\newcommand{\remark}[1]{}
 \remark{  }
\usepackage{bezier}
\usepackage{ifthen}
\begin{document}
\input{spllines}

\input{spldefs}

\title{The One Loop Effective Super-Potential and Non-Holomorphicity}
\author{{\bf Austin Pickering\thanks{e.mail:pickring@mth.kcl.ac.uk}\ \ and Peter West} \\ Dept. of Mathematics \\ King's College, London \\ WC2R 2LS}
\date{}
\maketitle
\setlength{\parindent}{2em}
\setlength{\unitlength}{0.0015 pt}
\thicklines

\remark{\input{splabs}}

\begin{abstract}
We calculate the \kayl\ and the lowest order non-\kayl\  contributions to the one loop effective superpotential using super-Feynman graphs in the massless \wzm, the massive \wzm\ and  \ymt.   We also calculate the \kayl\ term in  \nabymt\ for a general gauge group. Using this latter result we find the one loop \kayl\ contribution  for \ntwo\ \nabymt\ in terms of \none\ superfields and we show that it can only come from non-holomorphic contributions to the \ntwo\ effective potential.  
\end{abstract}

\vspace{10 ex}

\remark{\input{splintro}}

The effective potential, \biggam,  is the effective action with its external  momenta, $p_{i}$, set to
zero \bibnum{colewein}. For a supersymmetric theory we can calculate the one loop contributions of lowest order in the covariant derivatives  using the Feynman rules in $N=1$ superspace.
These are explained along with our conventions in \bib{westsusy} which also contains extensive references to the original literature.
For a generic $N=1$ supersymmetric theory with chiral superfields, \fye, and  gauge fields, \veee,
the effective potential   can be written as
\beq
\int \dz{8} \kay (\fyebar, \fye, V) + \eff{1} (\fyebar, \fye, V) \, \blockphi \nonumber \\ \mbox{} + \eff{2} (\fyebar, \fye, V) \, \upcov{A} \upcovbar{B} \veee \, \cov{A} \covbar{B} \veee + \ldots +
O(D^{4},\bar{D}^{4}) \elab{effpot}
\eeq
The first term, \kay,  in \eqn{effpot}  is referred to as the \kayl\ term. The first order non-\kayl\ terms
contain exactly two \cov{A} and two \covbar{B} derivatives and the \ldots indicate that we have not 
shown the terms where covariant derivatives act on both \veee\ and either \fye\ or \fyebar. 

To illustrate our techniques we first calculate the \kayl\ term and the lowest order non-\kayl\ term,
\eff{1}, for  the massless \wzm, which has no \veee\ fields, and then show how to extend our  results to the massive \wzm. The \kayl\ terms were found by a different method in \bib{buchbinder}.  In this method the superfields $\phi$ and
$\bar{\phi}$ are expanded about their background values and the superpropagator is found for the quantum perturbations.
The one loop effective potential is expressed in terms of this superpropagator which is  in turn
expressed in terms of a heat kernel by the Schwinger-de-Witt representation.
This results in a system of coupled differential equations which can be
solved to give an expression for the effective potential. Enforcing the restriction $\cov{A} \fye\ = 0$ gives the \kayl\ term.
The authors go on to give the form of the second term in the expansion, 
\eff{1}, but they do not calculate its coefficient explicitly. 
We shall show how this can be done using supergraphs.

Following this we apply similar methods to  \ymt. In this case  there is  
also an \eff{2} term and  $W_{A} = g \covbarsqu \cov{A} 
\veee$,  so we can get the \eff{2} \wsquind\ term from this. In  \nabymt\ we  calculate  the \kayl\ term for a general gauge group.
When the gauge group is in the adjoint
representation this theory becomes $N=2$ Yang-Mills theory without hypermultiplets. We show that the one-loop corrections we calculate in $N=1$ superspace
must come from a {\em non-holomorphic}  function of the chiral $N=2$ superfield $W$ and its
 conjugate $\bar{W}$.

\remark{\input{splmless}}

\vspace{4 ex}

\begin{figure}[ht]
\centering
\input{splpc01}
\caption{Order $n$ contribution to the one loop effective potential for the massless chiral field.}
\label{fig:mlessordnloop}
\end{figure}
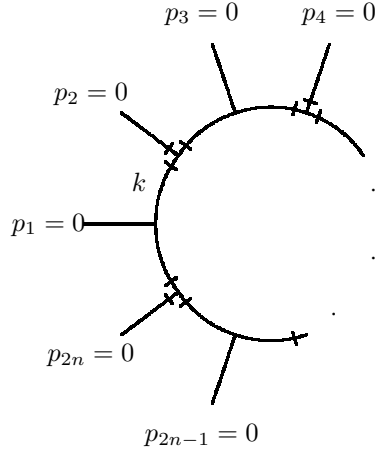

Our first example is the \wzm\ with $N=1$  chiral superfield, $\phi$, and the most general renormalisable action for this case 
 is :
\beq
S[\phi, \bar{\phi} ] &=& \int d^{8} z \bar{\phi} \phi + \int d^{6} z   \frac{m \phi^{2}}{2} + \frac{\lambda \phi^{3}}{3!} + \makebox{h.c.}.
\eeq
The effective potential is  expressed in terms of the  superfield  \fye\ and to begin with we set $m=0$.
The two vertices we have are $\lambda \fye^{3}/3!$ and its conjugate $\lambda \fyebar^{3}/3!$.
In the massless case we have only the \bfprop\ propagator and  consequently the two types of vertex  must be placed alternately around the loop, giving  $2n$ vertices in each loop, $n$ of each type, as shown in \fig{mlessordnloop}. 

In superspace the  expression for such a graph is 
\beq
\biggamord{n} &=& \int  \frac{ \df{\spct} \df{k} \dfs{\theta}{1} \ldots \dfs{\theta}{2n}}{ 2 n (16k^{2})^{2n}  }  \twoverts{1}{2}{3}   \twoverts{3}{4}{5}  \nonumber  \\ & &  \ldots \twoverts{(2n - 1)}{(2n)}{1}
\label{eqn:mlessordnloop}
\eeq
and by rearranging the covariant derivatives we can perform all but two of the \tht{i} integrations.
In \eqn{mlessordnloop}  we have included the combinatoric factor, $1/2n$, which is related to the symmetry of the loop \bibnum{colewein}, in this case $n$ for rotation and $n$ for reflection. Although the calculation in \eqn{mlessordnloop}  has been done for the massless \wzm\ the same procedure can be used for all the theories considered in this paper, giving the general expression 
\beq
\biggamord{n} \, = \,  \frac{1}{sn} \int \frac{\df{\spct} \df{k} \dfs{\theta}{1} \dfs{\theta}{2}}{16 \pi^{4}} \delta_{12} \extgen \dub ( \extgen \dub ( \extgen \ldots \dub \delta_{12} \ldots )) \elab{genfinal}
\eeq
where we have generalised the combinatoric factor to $1/sn$ where $s$ is an integer determined by  the symmetry of the loop. In this paper it is  2 for the \wzm\ and 1 for the gauge theories. In the theories we study in this paper we can combine the vertices into one effective vertex which we have called \extgen. We then connect $n$ copies of \extgen\ together to form the most general loop. These loops are then summed over $n$ from 1 to $\infty$ to give \biggam.
From \eqn{mlessordnloop} we see that in the massless \wzm\ 
\[
\extgen = \frac{\lambda^{2} \bar{\Phi} \Phi }{ 16 \ksqu^{2}}
\]
where $\ksqu = k^{2}$.

We can expand out the derivatives in \eqn{genfinal} using Leibniz' rule and then use the well known \delfn\ rules in superspace to deduce that only terms with equal even multiples of the superspace covariant derivatives acting on the external lines can contribute.  

Returning to the massless \wzm\ we see that, since $\Phi$ is chiral,
 we need not consider terms of the form
\(
\covbar{B} \cov{A} \fye
\)
because they are zero when the external momenta, $p_{i}$ are zero. This  enables us to rearrange derivatives without getting factors dependent on the $p_{i}$.
We can calculate the \kayl\ term, \kay, by not alowing any of the covariant derivatives to act on the external lines, \extgen, and then summing the graphs for all $n$. We write $\int \df{k} = \int  \pi^{2} \ksqu d \ksqu$ and the general expression is  
\beq
\kay & = & \int \frac{d \ksqu \df{\theta} }{16 s \pi^2 } \ln \left\{ 1 + 16 \extgen \ksqu \right\}. \elab{generalkay}
\eeq
For the massless \wzm\ this gives
\beq
\kay & = & \int \frac{d\ksqu \df{\theta} }{32 \pi^2 } \ln \left\{ 1 + \frac{\lambda^{2} \fyebar \fye}{\ksqu} \right\}.
\eeq

After  integrating  over $\ksqu$ we  impose wave-function renormalisation. 
Coleman and Weinberg \bibnum{colewein} give the standard renormalisation conditions for a non-supersymmetric  massive scalar field theory and discuss the necessary modifications when the mass in the original Lagrangian is zero. In a massive theory we can impose renormalisation conditions at  $\fye = \fye_{0} = 0$ but in the massless case  there is a divergent  logarithm  at this point and so we must renormalise at   $\fye = \fye_{0} \neq 0$ to avoid this.
Hence we demand \bibnum{buchbinder}
\beq
\renlims{\dtwogamma{\biggam} } & = &  1. \elab{rencondit}
\eeq
which leads to  our final result 
\beq
\kay {}_{\mathit{ren}} & = & \int \frac{ \dz{8} }{32 \pi^{2}} \lambda^{2} \fyebar \fye \left[ 2 - \ln \left\{ \frac{\fyebar \fye}{\mu^{2}} \right\} \right] \label{eqn:mlessgamoneren} .
\eeq
In \bib{buchbinder} the same result is achieved, with $\mu^{2} = \fyebar_{0} \fye_{0}$, using functional and operatorial methods in superspace instead of the more direct approach of supergraphs.

We move on  to the calculation of  terms in the expansion of \eqn{genfinal} with exactly two \cov{A} and two \covbar{B} acting on the external lines, \extgen.
As an example of how this could occur in the \supth{n} order term let the first `$p-1$' \dub\  
operators in \eqn{genfinal} act on the \delfn\ to give
\[
  \int \frac{\df{\spct} \df{k} \dfs{\theta}{1} \dfs{\theta}{2}}{
16 \pi^{4} s n}  \delta_{2  1}  \genvert{\extgen} (  \nonumber \\
\genvert{\extgen}  (  \ldots \dub\ ( \extgen^{p} \left( \dub\ \right)^{p} \delta_{1 2}) \ldots )
 ).
\]
By hypothesis the next \dub\ has at least one derivative acting on a \extgen\ term. One possibility is that {\em all four } of these derivatives act on the $\extgen^{p}$ term and so all the remaining \dub\ must act on the $\dub \delta$,  giving  
\[
  \int \frac{\df{\spct} \df{k} \dfs{\theta}{1} \dfs{\theta}{2}}{
16 \pi^{4} s n }  \genvert{\extgen^{n-p}} ( \extgen^{p} ) \, \delta_{2  1} \left( \dub\ \right)^{n-1} (\delta_{1
2}).
\]
where the $\theta_{2}$ integral is now trivial. We could allow other combinations of covariant derivatives to act on the $\extgen^{p}$ term and the remaining derivatives from the \dub\ would act on the final \delfn. In this case we would need to get the remainder of our required four derivatives from subsequent \dub\ operators, parts of which act on the external lines, \extgen, and parts of which act on the \delfn\ by Leibniz rule.
 In total there are nine ways of getting four derivatives to act on the external lines and they are displayed in  \tab{tab1} where  $\sumfac{n} =  (-16 \ksqu )^{n} / n\ksqu $ and the $c_{i}$ are combinatoric factors needed to combine the results into the final answer. 
\begin{table}[ht]
\begin{center}
\input{spltb01}
\end{center}
\kap{Contributions to the four derivative term.}
\tlab{tab1}
\end{table}
From this table we find, after some algebra, that for a general \extgen\
\beq
\sum_{i=1}^{9} c_{i} I_{i} &=& \frac{- \, (\upcov{A} \extgen \upcovbar{B} \extgen -  (1+\extgen) \upcov{A} \upcovbar{B} \extgen ) \, (\cov{A} \extgen \covbar{B} \extgen 
 - (1+\extgen) \cov{A} \covbar{B} \extgen )}{2 \ksqu (1+\extgen)^{4}}    \elab{simplextdrv}
 \eeq       
In the massless \wzm\ \extgen\ corresponds to  $\lambda^2 \fyebar \fye / ( 16 \ksqu^{2} )$ and after simplification and integration over $\ksqu$ we find
\beq
\eff{1} \; = \;  \int \frac{d\ksqu \dz{8} }{(16 \pi)^{2} s} \sum_{i=1}^{9} c_{i} I_{i}  &=& \int \frac{ \dz{8} }{(4 \pi)^{2}} \frac{\blockphi}{384 ( \fyebar \fye )^{2} }.
\eeq
The form of this result was given in \bib{buchbinder} but the coefficient is given only in the form of an integral, for which the result is not known. Our technique gives  the  coefficient explicitly and the 
 complete result for the massless \wzm\ is \beq
\biggam = \int \frac{ \dz{8} }{(4 \pi)^{2}} \frac{\lambda^{2} \fyebar \fye}{2} \left[ 2 - \ln \left\{ \frac{\fyebar \fye}{\mu^{2}} \right\} \right] +  \frac{(D \fye)^{2} (\bar{D} \fyebar)^{2}}{384 ( \fyebar \fye )^{2} } + O(D^{4}, \bar{D}^{4})
\eeq
with $(D \fye)^{2} (\bar{D} \fyebar)^{2} = \blockphi$.

\remark{\input{splmass}}

\vspace{4 ex}

When we consider the massive \wzm\  we can no longer neglect graphs containing adjacent
identical vertices as we now have \ffprop\ and \bbprop\ propagators aswell as \bfprop. To account
for this we replace each vertex in the massless diagrams by $q+1$ adjacent copies of this vertex and
sum $q$ from 0 to $\infty$ at each point.  This corresponds to defining a dressed vertex so that the
chiral vertex becomes
\beq
\lambda \fye \covbarsqu \sum_{q = 0}^{\infty} \left(  \frac{- m \lambda \fye}{ k^{2} + m^{2}  }
\right)^{q} & = & \frac{\lambda \fye }{1 + \frac{m \lambda \fye}{ k^{2} + m^{2}  } } \covbarsqu
\eeq
and by using it we allow for all possible numbers of adjacent chiral vertices from one to infinity
between any two neighbouring anti-chiral vertices.
The resulting contribution  behaves just like a single chiral vertex as far as the \dalg\ is concerned
and after similarly dressing the  anti-chiral vertex we can only connect the dressed vertices using
the \bfprop\ propagator as we have allowed for all possible occurrences of the other two
propagators.
What is more, the symmetry arguments for the combinatoric factors will follow through and once
again $s = 2$. \remark{because whilst any particular subgraph contained in the \supth{n} order
dressed graph  may well not rotate (reflect) into itself in $n$ $( n )$ ways it does rotate (reflect)
into another member of the multi-infinite collection of graphs represented by the \emph{same
\supth{n} order dressed graph} in exactly $2n$ ways. }
The situation is very similar to the massless case but we have a new expression for the external
lines, \extgen\ given  by \[ \extgen = \frac{\lambda^{2} \fyebar \fye  }{ 16 (\ksqu+ m \psye)(\ksqu + m
\psyebar)} \]
where
$\psye = \lambda \fye + m$. After integrating and renormalising  using \eqn{rencondit}, we get
\beq
\kay  {}_{\mathit{ren}} &=& \int \frac{\dz{8} \psyebar \psye }{32 \pi^{2}} \left[ 2 - \ln \left(
\frac{\psyebar \psye}{\mu^{2}} \right) \right].
\eeq

The same arguments about dressing vertices give
\beq
\eff{1} &=& \int \frac{ \dz{8} }{(4 \pi)^{2}} \frac{\blockpsye}{384 ( \psyebar \psye )^{2} }
\eeq
for the massive non-\kayl\ term.
The complete result is thus
\beq
\biggam = \int \frac{ \dz{8} }{(4 \pi)^{2}} \frac{\psyebar \psye }{2} \left[ 2 - \ln \left(
\frac{\psyebar \psye}{\mu^{2}} \right) \right] + \frac{(D \psye)^{2} (\bar{D} \psyebar)^{2}}{384 (
\psyebar \psye )^{2} } + O(D^{4}, \bar{D}^{4})
\eeq
These results agree with those found by the method of \bib{buchbinder} if  $\mu^{2} =
\psyebar_{0} \psye_{0}$ and we take the value of the unknown coefficient to have the value given.

\remark{\input{spluone}}

\setcounter{gdots}{10}

\vspace{ 4 ex}

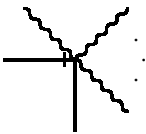
\begin{figure}[ht]
\centering
\begin{picture}(160000, 24000)(-80000,-12000)
\put(0,0){ \gline{-18}{18}
\gline{18}{18}
\gline{18}{-18}
\sline{0}{-25} \mline{-25}{0}
}
\put(15500, 5000){\circle*{1000}}
\put(17500, 0){\circle*{1000}}
\put(15500,-5000){\circle*{1000}}
\end{picture}
\caption{The general form of an interaction vertex from the $\fyebar e^{g V} \fye $ term in \ymt. }
\label{fig:genymvert}
\end{figure}

We move on to the case of  \ymt. The action is
\beq
\frac{1}{64 g^{2}} \int \mathrm{d}^{6} z \; \wsquind + \int \mathrm{d}^{8} z \; \fyebar e^{g V} \fye +
\mathrm{gauge \; fixing + ghosts} \elab{ymact}
\eeq
where $W_{A} = \covbarsqu e^{-gV} e^{gV} \cov{A} V$ so that for a U(1) gauge group $W_{A} = g \covbarsqu  \cov{A} V$.
We have two propagators, \bfprop\ and \vvprop\ and  the interactions come from
the $\fyebar e^{g V} \fye$ term. They give
vertices with any number of $V$ lines and one \fyebar\ and one \fye\ line, as indicated in \fig{genymvert}.

\begin{figure}[ht]
\centering
(a)
\begin{picture}(45000,18000)(-20000,0)
\put(0,0){ \mline{-18}{18} \gline{0}{25} \sline{18}{18} \gline{-25}{0} \gline{25}{0} }
\end{picture}
(b)
\begin{picture}(45000,18000)(-20000,0)
\put(0,0){ \gline{-18}{18} \mline{-25}{0} \sline{18}{18} \gline{25}{0}  }
\end{picture}
(c)
\begin{picture}(45000,18000)(-20000,0)
\put(0,0){ \gline{-18}{18} \sline{-25}{0} \mline{18}{18} \gline{25}{0}  }
\end{picture}
(d)
\begin{picture}(40000,18000)(-20000,0)
\put(0,0){  \mline{25}{0} \sline{-25}{0} \gline{0}{25} }
\end{picture}
\caption{Possible ways of orienting vertices in a loop in \ymt. }
\label{fig:ymvertopts}
\end{figure}
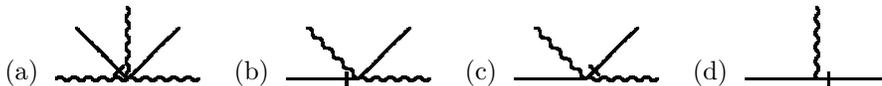

Exactly two of these lines must be internal giving the options shown in \fig{ymvertopts} where
horizontal lines are internal, the $V$ lines are the wiggly lines and although we have only shown only one external $V$ line in each case there could infact be any number, from zero to infinity, for each graph. However \ymvrt{d} requires at least one external $V$ line to be non-zero.

As we have seen the key to the solution is to find a general expression for all the allowed loops
and cast it in the form of \eqn{genfinal}. To do this  we observe that the vertex in \ymvrt{b} cannot
use the \vvprop\ propagator to join to another identical vertex in the \kayl\ approximation since
$\covsqu \covsqu = 0$. Similarly for \ymvrt{c} and so any loop with \ymvrt{b} in has an equivalent
number of \ymvrt{c} vertices. In fact they go in pairs joined by their \vvprop\ propagator but this
propagator can be dressed with any number of copies of \ymvrt{a} as shown in \fig{ymcombvert}.
\bef[h]
\beq
\sum_{n = 0}^{\infty} \bep(40000,10000)(-20000,8000) \put(0,0){\sline{-25}{0} \mline{0}{25}
\gline{25}{0} } \eep \left( \bep(34000,12000)(-14000,9000) \put(0,0){\sline{-18}{18} \gline{0}{25}
\gline{25}{0} \mline{18}{18} } \eep \right)^{n}\bep(20000,10000)(-1000,8000)
\put(0,0){\mline{25}{0} \sline{0}{25} } \eep & = & \begin{picture}(60000, 10000)(-30000,8000)
\put(-9380,0){ \gline{-18}{18}
\mline{0}{25}
\gline{25}{0} \sline{-25}{0}
}
\put(9380,0){\sline{0}{25}
\gline{18}{18}
\mline{25}{0}
}
\put(0,0){\circle{6000}}
\end{picture}
\eeq
\kap{The dressed gauge propagator.}
\flab{ymcombvert}
\eef

We could form a valid loop by connecting $n$ copies of this vertex together, but because \ymvrt{d} contributes the same internal lines to the loop, it is equally valid to  form a new loop by substituting \fig{ymcombvert} with \ymvrt{d}  at any point in the loop . This corresponds to creating a combined vertex
by adding these two together as in \fig{ymtotalvert}. {
\bef[h]
\[
\raisebox{5000\unitlength}[15000\unitlength][5000\unitlength]{
\begin{picture}(40000,18000)(-20000,9000)
\put(0,0){  \mline{25}{0} \sline{-25}{0} \gline{0}{25} }
\end{picture}
+ \begin{picture}(60000, 18000)(-30000,9000)
\put(-9380,0){ \gline{-18}{18} \mline{0}{25} \gline{25}{0} \sline{-25}{0} }
\put(9380,0){\sline{0}{25}
\gline{18}{18}
\mline{25}{0}
}
\put(0,0){\circle{6000}}
\end{picture}
}
\]
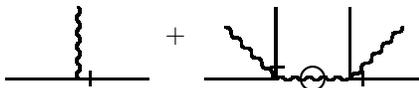
\kap{The combined vertex from which all one loop graphs can constructed in \ymt.}
\flab{ymtotalvert}
\eef
}
Joining $n$  such a vertices  together   and performing the $\sum_{n=1}^{\infty}$ gives all
contributions to the effective superpotential in the form of \eqn{genfinal}. By this procedure we have neglected only
those graphs which consist of \ymvrt{a}. These are identically zero because they contain no \dub\
between the \delfn s.

Using the super-Feynman rules and Dyson's formula for the combinatoric factor of a tree diagram
we find that
\beq
\extgen &=& \frac{1 }{16 \ksqu}  \left\{ e^{g V} - 1 + \frac{\beta \fyebar e^{2g V} \fye}{1 - \beta
\fyebar e^{g V} \fye} \right\}  ,  \elab{uonetoteffvert}
\eeq
with     $\beta = \frac{- g^{2}}{\ksqu}$.
There is no reflection symmetry in this case so $s = 1$ in \eqn{genfinal} and we get
\beq
\kay & = &  \int \frac{d\ksqu \dz{8}}{16 \pi^{2}} \left\{ g V - \ln \left( 1 + \frac{g^{2} \fyebar e^{g
V} \fye}{\ksqu} \right) \right\} . \elab{cfkayl}
\eeq

We can ignore the $g V$ term as it would not appear in a dimensional reduction
scheme \bibnum{superspace}. We integrate and then  renormalise using \eqn{rencondit}
where $\fye_{0} \neq 0$ because \fye\ is massless and we must set $V = 0$ at the renormalisation point. We get
\beq
\kay & = & - \int \frac{\dz{8}}{(4 \pi)^{2}} g^{2} \fyebar e^{g V} \fye \left\{ 2 - \ln \left(
\frac{\fyebar e^{g
V} \fye}{\mu^{2}} \right) \right\}. \elab{abkayl}
\eeq
where $\mu^{2} = \fyebar_{0} \fye_{0}$.

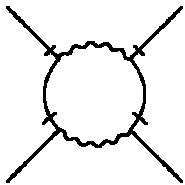
\begin{figure}[ht]
\centering
\input{splpc99}
\caption{Loop containing adjacent \covsqu\ which can only contribute in the non-\kayl\ case.}
\label{fig:extragrps}
\end{figure}

We can use \eqn{uonetoteffvert} to calculate some contributions to 
the non-\kayl\ term aswell, but because we are no longer in the \kayl\    
approximation we cannot simply ignore graphs with adjacent \covsqu\ operators 
as we did in the \kayl\ case.  Hence we must also include the graph shown in 
\fig{extragrps}. In order that this is non-zero one of the adjacent \covsqu\ must act on the external lines, or else we get 
\covsqu \covsqu\ which is zero. Similarly for the adjacent \covbarsqu, and thus we immediately get the required number
of external covariant derivatives. It remains to dress  the 
propagators 
with all allowed vertices, without letting any derivatives act externally.
For a U(1) gauge group the only terms we find are infrared divergent. It can be shown that in a full 
treatment with massive scalars  these terms have  appropriate
factors of $1/m^{2}$ associated with them.     Hence the only contributions
to the non-\kayl\ term come from substituting the expression for   \extgen\
into  \eqn{simplextdrv} as before.

We are interested in two terms in particular: the first where none of the derivatives act on external
$V$ lines, the second where they all do and we only include terms of $O(V^{2})$. In the former case we get
\beq
\eff{1} &=& \int \frac{ \dz{8} }{(4 \pi)^{2}} \frac{\blockphi}{192 ( \fyebar \fye )^{2} }
\eeq
whilst in the latter we get 
\beq
\eff{2} &=&  \frac{1}{16 g^{2}} \int \frac{ \dz{8} }{(4 \pi)^{2}} \frac{ g^{2} \upcov{A}
\upcovbar{B} V \cov{A} \covbar{B} V}{2} \left\{ \ln \left( \frac{\fyebar  \fye}{\mu^{2}} \right)  \right\}. \elab{abwsqu}
\eeq
This is clearly supersymmetric but we can use $\int \dt{\bar{\theta}} = - \covbarsqu / 4$ to write this in terms of \wsquind. In order to show that \eqn{abwsqu} is gauge invariant we must use the fact that the external momenta are zero. Finally, in order to renormalise, we have set the coefficient of $W^{A} W_{A}$ to be $1/64g^{2}$ at $\fye = \fye_{0}$. 
The full result is thus
\beq
\biggam = \int \frac{\df{\theta}}{(4 \pi)^{2}} g^{2} \fyebar e^{g V} \fye \left\{ \ln \left(
\frac{\fyebar
\fye}{\fyebar_{0} \fye_{0}} \right) - 2 \right\} + \frac{\blockphi}{192 ( \fyebar \fye )^{2}} - \nonumber
\\ \frac{1}{64 g^{2}} \int \frac{ \dt{\theta} }{(4 \pi)^{2}} \frac{ g^{2} W^{A} W_{A}}{2} \left\{ \ln \left(
\frac{\fyebar  \fye}{\mu^{2}} \right)  \right\} + \ldots  + O(D^{4}, \bar{D}^{4})
\eeq
where the \ldots denote terms of $O(D^{2}, \bar{D}^{2})$ which do not interest us.

\remark{\input{splnab}}

\vspace{4 ex}

\bef[ht]
\centering
\begin{picture}(40000, 30000)(-20000,-5000)
\put(0,0){\mline{-10}{23} \gline{25}{0} \gline{-25}{0} \sline{10}{23} }
\put(-11000, 20000){\makebox(0,0){$\fyebar^{c}$}}
\put(11000, 20000){\makebox(0,0){$\fye_{d}$}}
\put(8000, -3000){\makebox(0,0){b}}
\put(-8000, -3000){\makebox(0,0){a}}
\end{picture}
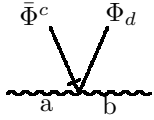
\kap{The non-\abe\ vertex used to dress the gauge propagator.}
\flab{nabdressvert}
\eef

We can extend our \uone\ arguments to the  case of \nabymt\ with a general gauge group, but we must obtain a new
expression for the combined vertex in \fig{ymtotalvert}. For simplicity we shall set all external $V$
lines to zero.
The vertex shown in \ymvrt{a} becomes that shown in \fig{nabdressvert}.
Hence we define 
\beq
\mat{A}{a}{b} = \frac{-g^{2}}{2} \fyebar \left\{ \tee{a}, \tee{b} \right\} \fye.
\eeq
In the absence of external lines the combined vertex becomes 
\beq
{}_{a} {\begin{picture}(60000, 20000)(-22000,0)
\put(0,0){\mline{-25}{0} \sline{0}{25} \gline{25}{0} }
\put(18760,0){\sline{25}{0} \mline{0}{25}}
\put(9380,0){\circle{6000}}
\end{picture}}_{b}
&=& - g^{2} \fulltee{c}{a}{e} \fye_{e} \mat{(\frac{1}{\ksqu-A})}{c}{d} \fyebar^{f} \fulltee{d}{f}{b}  \left( \frac{\covbarsqu \covsqu}{16 \ksqu} \right). \elab{noextvee}
\eeq
Hence \[ \extgen = - g^{2}  \fulltee{c}{a}{e} \fye_{e} \mat{(\frac{1}{16r(\ksqu-A)})}{c}{d} \fyebar^{f} \fulltee{d}{f}{b}  \] and
\beq
\kay &=& \int \frac{d\ksqu \dz{8}}{16 \pi^{2}} \left[ \trce \ln  \left(\ksqu - F \right) - \trce \ln  \left( \ksqu -
A \right) \right] \elab{lamzero}
\eeq
where 
\beq
F = \frac{g^{2}}{2} \fyebar \left[ \tee{a}, \tee{b} \right] \fye.
\eeq

We can perform the integral and by extending \eqn{rencondit} to matrix form we eventually get
\beq
\kay &=& \trce \int \frac{\dz{8}}{16 \pi^{2}} \left[ \Delta - \frac{\fyebar \fye}{\fyebar_{0}
\fye_{0}} {\fyebar_{0}}^{a} \renlims{\dtwogammaind{\Delta}{a}{b}} {\fye_{0}}_{b} \right]. \elab{ethr}
\eeq
where
$\Delta = F \ln |F| - A \ln |A|$, i.e.
\beq
\mat{\Delta}{a}{c} &=& \frac{g^{2}}{2} \fyebar \left[ \tee{a} ,\tee{b} \right] \fye \mat{\left( \ln  \left|
\frac{
g^{2}}{2} \fyebar \left[ T ,T \right] \fye \right| \right)}{b}{c} \nonumber \\ & & \mbox{} +
\frac{g^{2}}{2}
\fyebar \left\{ \tee{a} ,\tee{b} \right\} \fye \mat{\left( \ln \left| \frac{g^{2}}{2} \fyebar \left\{ T,T
\right\} \fye \right| \right) }{b}{c} . \elab{deldef}
\eeq
Because all the matrices we have to deal with are linear in each of \fyebar\ and \fye\ we can infact
simplify this expression further, giving the general result
\beq
\kay &=&  \int \frac{\dz{8}}{16 \pi^{2}} \left( \alpha_{0}  \fyebar \fye  + \trce \Delta
\right). \elab{gennabfin}
\eeq
where
\beq
\alpha_{0} &=& - g^{2} C(R) - \frac{\Delta_{0}}{\fyebar_{0} \fye_{0}}
\eeq
and 
\[
{(T^{s})_{a}}^{b} {(T^{s})_{b}}^{c} = {\delta_{a}}^{b} C(R).
\]
We have
 investigated the form of the \wsquind\ term in the non-\abe\ case and we shall publish our results
in a forthcoming paper.

We now evaluate \eqn{gennabfin} when the  gauge group is  SU(2) in the adjoint representation.
In this case the theory possesses $N=2$ supersymmetry and \[
\fulltee{a}{b}{c} = \epsilon_{a b c}
\] the totally anti-symmetric tensor in three dimensions. We can diagonalise the matrices in
\eqn{gennabfin} and the trace becomes a sum over { \em non-zero} eigenvalues. To prove this we must diagonalise before we integrate over $\ksqu$.
For $A$ the eigenvalues are \[ -a, \;  -d_{+}, \; -d_{-} \; \; \;  \mathrm{where} \; \; \;
d_{\pm} = \frac{-g^{2}}{2} (\fyebar \fye \pm \sqrt{\fyebar^{2} \fye^{2}} ) \] \remark{for $B$ they are \[ 0, \; b, \; b \; \; \;  \mathrm{where} \; \; \;  b =
g^{2} \fyebar \fye \]} and for $F$ they are \[ 0, \;  \pm f \; \; \;   \mathrm{where} \; \; \;  f = \frac{g^{2}}{2} \sqrt{(\fyebar
\fye)^{2} - \fyebar^{2} \fye^{2}}. \]
Hence
\remark{\( \trce \Delta =
\)
and we have }
\beq
\kay = \int \frac{\df{\theta}}{16 \pi^{2}} \alpha_{0}  \fyebar \fye + 2 a \ln a + d_{+} \ln
d_{+} + d_{-} \ln d_{-}. \elab{kadj}
\eeq

\remark{\input{splconc}}

\vspace{ 4 ex}

In this paper we have calculated our results for $N=1$ supersymmetric theories. \nabymt\ also possesses $N=2$ supersymmetry when the matter fields are in the adjoint representation of the gauge group. For the \none\ action we have considered the resulting theory is \ntwo\ \nabymt\ without hypermultiplets. By combining the two $N=1$ fields \fye\ and $W_{A}$ into a single chiral $N=2$ superfield $W$ we can write the action in $N=2$ superspace. 
It was shown in  \bib{seib}  that a holomorphic function $\mathcal{F}$ of the $N=2$ Yang-Mills
chiral superfield $W$ can be written in $N=1$ language as
\beq
\int \df{\theta} \mathrm{Im} \left( \fyebar \frac{\partial \mathcal{F}}{\partial \fye} \right) + \frac{1}{2}
\int d^{2} \theta \wsquind \mathrm{Im} \left( \frac{\partial^{2} \mathcal{F}}{\partial \fye^{2} } \right).
\eeq
Thus if there are holomorphic corrections to the one loop effective potential in $N=2$ Yang-Mills
theory the $N=1$ \kayl\ term should be related to the holomorphic function $\mathcal{F}$ as
above. In order for this to be so we need to  use a relation analagous to\[
\ln \fyebar \fye = \ln \fyebar + \ln \fye, \] enabling us to split the logarithm in \eqn{ethr} into a sum of chiral and anti-chiral parts. There is no such relation for  $\ln \fyebar\{\tee{a},
\tee{b}\} \fye$ or $\ln \fyebar \left[ \tee{a},
\tee{b}\right] \fye$ and  so these terms must
come from a {\em non-holomorphic} function of $W$ and $\bar{W}$. 

\vspace{ 4 ex}

In this paper we have  applied the supergraph technique to the \wzm\ to calculate both the \kayl\ contribution  and the first non-\kayl\ contribution  to the one loop effective superpotential.    We have extended these methods  to  \ymt\ where we use our non-\kayl\
calculations to find not only the term in \blockphi\ but also the \wsquind\ term.
In addition we have also calculated the \kayl\ term in \nabymt\ for a general gauge group
and and we conclude that it can only be generated from a non-holomorphic correction to the
effective potential in $N=2$ Yang-Mills theory.

\vspace{4 ex}
\setlength{\parindent}{0 em}

\bnt{Acknowledgement: } We would especially like to thank P. di Vecchia, I.  Buchbinder and J. Yarevskaya for useful
discussions. This work was undertaken as part of an
E.P.S.R.C. funded PhD. studentship.
\ent

\vspace{4 ex}

\bnt{Note Added: } After we had finished this work  we became aware of a paper \bibnum{wgr} in
which the authors calculate the one loop \kayl\ correction to effective super-potential for $N=2$ Yang-Mills Theory. Like us they find that there are
non-holomorphic corrections to the $N=2$ Yang-Mills effective potential and whilst they do not
calculate any non-\kayl\ terms they do calculate the function of the $N=2$ fields
which gives rise to  these non-holomorphic corrections.
\ent

\vspace{4 ex}

\bnt{Note Added In Proof : } 
In the above we have found the \kayl\ contribution to the effective superpotential for \nabymt\ in the absence of cubic interaction terms.
In this note we extend our result slightly by adding a term of the form 
\[
\int d^{2} \theta \frac{1}{3!} \lambda^{abc} \fye_{a} \fye_{b} \fye_{c}  +     \mbox{ h.c.}
\] to the  action of \eqn{ymact}, and we obtain the one loop \kayl\ contribution to the superpotential for the most general renormalisable $N=1$ supersymmetric theory.
We find the \kayl\ contribution to the superpotential by considering all graphs of the form shown in \fig{mlessordnloop} and then dressing the \fbprop\ propagator with the vertex shown in \eqn{noextvee}. Similarly the \bfprop\ is dressed with the transpose of this vertex. We use \eqn{noextvee} to define $C$ as 
\beq
{}_{a} {\begin{picture}(58000, 20000)(-21000,0)
\put(0,0){\sline{-25}{0} \mline{0}{25} \gline{25}{0} }
\put(18760,0){\mline{25}{0} \sline{0}{25}}
\put(9380,0){\circle{6000}}
\end{picture}}_{b}
&=& - g^{2} \fulltee{c}{a}{e} \fye_{e} \mat{(\frac{1}{\ksqu-A})}{c}{d} \fyebar^{f} \fulltee{d}{f}{b}  \left( \frac{\covbarsqu \covsqu}{16 \ksqu} \right) \; \; = \; \;  {C_{a}}^{b}  \left( \frac{\covbarsqu \covsqu}{16 \ksqu} \right). 
\eeq
The dressed matter  propagators are then given by  
\beq
\fbprop_{\mathrm{dressed}} \; \; = \; \; 
\begin{picture}(60000, 10000)(-22000,0)
\put(0,0){ \rline{35}{0} }
\put(13132,0){\circle{6000}}
\end{picture} &=& \frac{1}{r} \sum_{n=0}^{\infty} (-C)^{n} \; \; = \; \; \frac{1}{r} \frac{1}{1+C}
\\
\bfprop_{\mathrm{dressed}} \; \; = \; \; 
\begin{picture}(60000, 10000)(-22000,0)
\put(0,0){ \mline{35}{0} }
\put(13132,0){\circle*{6000}}
\end{picture} &=& \frac{1}{r} \sum_{n=0}^{\infty} (-C^{T})^{n} \; \; = \; \; \frac{1}{r} \frac{1}{1+C^{T}}
\eeq
We can generate all possible graphs containing at least one $\lambda \Phi^{3} / 3!$ vertex, by forming a loop with $n$ of the  vertices shown in \eqn{drsvrt}. 
\beq
{}_{c} {\begin{picture}(48000, 20000)(-11000,0)
\put(0,0){\sline{-15}{0} \sline{0}{25} \rline{26}{0} }
\put(18760,0){\mline{25}{0} \mline{0}{25}}
\put(9380,0){\circle{5000}}
\put(28140,0){\circle*{5000}}
\end{picture}}_{g}
 \;  =  \;  {\extgen^{c}}_{g} \covbarsqu \covsqu \;  =  \; (\lambda^{abc} \fye_{a}) {\left( \frac{1}{1+C} \right)_{b}}^{d}
 (\lambda_{edf} \fyebar^{e}) {\left( \frac{1}{1+C^{T}} \right)^{f}}_{g} \left( \frac{\covbarsqu \covsqu}{16 \ksqu^{2}} \right) \elab{drsvrt}
\eeq
Substituting this expression for \extgen\ into \eqn{generalkay} gives the corresponding contribution to the superpotential.
\ent
\setlength{\parindent}{2em}

In addition to this we must include the contribution from all loops without any $\lambda \Phi^{3} / 3!$ vertices, but this is simply the result obtained in \eqn{lamzero} where $\lambda^{abc} = 0$. Rewriting this in terms of $C$ and adding these two contributions together gives the full massless result
\beq
\kay &=& \int \frac{\dz{8} d \ksqu }{32 \pi^{2}}  \left[   \trce \ln  \left( 1 +  \frac{1}{\ksqu}  P \left( \frac{1}{1+C} \right) \bar{P} \left( \frac{1}{1+C^{T}} \right) \right) + 2 \trce \ln  \left( 1 + C \right) \right]  
\eeq 
where $P^{bc} = \lambda^{abc} \fye_{a}$. 

We can now deduce the result for the massive theory by treating the mass term, $m^{ab} \fye_{a} \fye_{b} / 2!$, as an interaction vertex. This vertex behaves in exactly the same way as the $\lambda \Phi^{3} / 3!$ vertex, but with $P^{bc}$ replaced by $m^{bc}$. We can thus simply add these vertices together, and the massive results then follow from the massless results under the transformation  
\beq 
P^{bc} \; \; \rightarrow \; \; \psye^{bc} &=& \lambda^{abc} \fye_{a} + m^{bc},
\eeq
as we found explicitly for the \wzm. 

We get 
\beq
\kay &=& \int \frac{\dz{8} d \ksqu }{32 \pi^{2}}  \left[   \trce \ln  \left( 1 +  \frac{1}{\ksqu}  \psye \left( \frac{1}{1+C} \right) \psyebar \left( \frac{1}{1+C^{T}} \right) \right) + 2 \trce \ln  \left( 1 + C \right) \right]  
\eeq 
which is the \kayl\ contribution for the most general massive $N=1$ renormalisable theory with cubic interactions and a general gauge group. 

After we had obtained these results we became aware of a recent paper, which appeared on the hep-th archive before this `note added in proof' was written, and which derived the results by a different method, for a general gauge choice. We have derived them for the super Fermi-Feynman gauge, $\xi = 1$ in their notation, as explained along with our conventions in \bib{westsusy}. Although the two expressions look different, after some algebra we find that they are the same, provided that the matrix \[ \left[ \frac{1}{k^{2} - (S + T)} \right] \] in Equation (4.17) of \bib{gru} is replaced by its transpose. 

In this paper \bibnum{gru} the authors also find an additional imaginary contribution to the \kayl\ term for SU(2) \nabymt. This term can be recovered from our results by removing the modulus signs in \eqn{deldef}. However, they also note that such a term does not contribute to the physics.

\remark{\input{splbib}}

\end{document}

%% file: spllines.tex
\newcounter{dots}
\setcounter{dots}{7}
\newcounter{gdots}
\setcounter{gdots}{4}
\newcounter{its}
\setcounter{its}{5}

\newcounter{gx}
\newcounter{gxa}
\newcounter{gxb}
\newcounter{gy}
\newcounter{gya}
\newcounter{gyb}
\newcounter{stepup}
\newcounter{gxc}
\newcounter{gyc}
\newcounter{gxd}
\newcounter{gyd}

\newcounter{mz}
\newcounter{mx}
\newcounter{mxa}
\newcounter{mxb}
\newcounter{myb}
\newcounter{mxc}
\newcounter{myc}
\newcounter{my}
\newcounter{mya}
\newcounter{mlb}
\newcounter{mla}
\newcounter{coun}
\newcounter{mv}
\newcounter{mdots}
\newcounter{fdots}

\setcounter{stepup}{6}

\newcommand{\newmatterline}[6]{
\setcounter{mdots}{#1}
\setcounter{mx}{#2}
\setcounter{my}{#3}
\setcounter{mz}{#4}
\setcounter{mv}{#5}
\setcounter{coun}{0}
\setcounter{mya}{0}
\setcounter{mxa}{0}
\setcounter{myb}{0}
\setcounter{mxb}{0}
\setcounter{fdots}{0}
\whiledo{#6 > \value{coun}}{\addtocounter{coun}{1} \addtocounter{fdots}{\value{mdots}}
\addtocounter{mxa}{\value{mx}} \addtocounter{mya}{\value{my}}
\addtocounter{mxb}{\value{mz}} \addtocounter{myb}{\value{mv}} }
\setcounter{my}{\value{mya}}
\setcounter{mx}{\value{mxa}}
\put(0,0){\begin{picture}(0,0)
\addtocounter{mxa}{\value{mx}}
\addtocounter{mya}{\value{my}}
\setcounter{my}{\value{mya}}
\setcounter{mx}{\value{mxa}}
\addtocounter{mxa}{\value{mx}}
\addtocounter{mya}{\value{my}}
\setcounter{mz}{\value{mxb}}
\setcounter{mv}{\value{myb}}
\addtocounter{mxb}{\value{mz}}
\addtocounter{myb}{\value{mv}}
\setcounter{mz}{\value{mxb}}
\setcounter{mv}{\value{myb}}
\addtocounter{mxb}{\value{mz}}
\addtocounter{myb}{\value{mv}}
\setcounter{mxc}{\value{mz}}
\setcounter{myc}{\value{mv}}
\addtocounter{mxb}{-\value{mv}}
\addtocounter{myb}{\value{mz}}
\addtocounter{mxc}{\value{mz}}
\addtocounter{myc}{\value{mv}}
\ifthenelse {\value{mx} = 0} { \ifthenelse{\value{my} > 0} { \put(0,0) {\line(0,1){\value{mya}}} }{
\put(0,0) {\line(0,-1){-\value{mya}}}  }} { \ifthenelse{\value{my} = 0} { \ifthenelse{\value{mx} > 0}
{ \put(0,0) {\line(1,0){\value{mxa}}} }{  \put(0,0) {\line(-1,0){-\value{mxa}}}  }} {
\put(0,0) {\bezier{\value{fdots}}(0,0)(\value{mx},\value{my})(\value{mxa},\value{mya})} } }
\put(\value{mxb},\value{myb}) {\bezier{\value{mdots}}(0,0)(\value{mv},-
\value{mz})(\value{myc},-\value{mxc})}
\end{picture}
}}


\newcounter{cx}
\newcounter{cxa}
\newcounter{cy}
\newcounter{cya}
\newcounter{cz}
\newcounter{cza}
\newcounter{cv}
\newcounter{cva}
\newcounter{disp}

\newcommand{\curvedmatterline}[7]{
\setcounter{mdots}{#1}
\setcounter{mx}{#2}
\setcounter{my}{#3}
\setcounter{mz}{#4}
\setcounter{mv}{#5}
\setcounter{coun}{0}
\setcounter{mya}{0}
\setcounter{mxa}{0}
\setcounter{cya}{0}
\setcounter{cxa}{0}
\setcounter{cza}{0}
\setcounter{cva}{0}
\setcounter{myb}{0}
\setcounter{mxb}{0}
\setcounter{fdots}{0}
\whiledo{#6 > \value{coun}}{\addtocounter{coun}{1} \addtocounter{fdots}{\value{mdots}}
\addtocounter{mxa}{\value{mx}} \addtocounter{mya}{\value{my}}
\addtocounter{mxb}{\value{mz}} \addtocounter{myb}{\value{mv}} }
\setcounter{coun}{0}
\setcounter{disp}{#7}
\ifthenelse{#7 < 0}{ \setcounter{cy}{-\value{cy}} \setcounter{cx}{-\value{cx}} \setcounter{cv}{-\value{cv}} \setcounter{cz}{-\value{cz}} \setcounter{disp}{-\value{disp}} }{}
\whiledo{\value{disp} > \value{coun}}{\addtocounter{coun}{1} \addtocounter{cxa}{-\value{cy}} \addtocounter{cya}{\value{cx}} \addtocounter{cva}{\value{cz}} \addtocounter{cza}{-\value{cv}}}
\addtocounter{mxb}{\value{cza}}
\addtocounter{myb}{\value{cva}}
\setcounter{cy}{\value{cya}}
\setcounter{cx}{\value{cxa}}
\addtocounter{cxa}{\value{cx}}
\addtocounter{cya}{\value{cy}}
\setcounter{my}{\value{mya}}
\setcounter{mx}{\value{mxa}}
\addtocounter{mxa}{\value{mx}}
\addtocounter{mya}{\value{my}}
\setcounter{my}{\value{mya}}
\setcounter{mx}{\value{mxa}}
\addtocounter{mxa}{\value{mx}}
\addtocounter{mya}{\value{my}}
\setcounter{mz}{\value{mxb}}
\setcounter{mv}{\value{myb}}
\addtocounter{mxb}{\value{mz}}
\addtocounter{myb}{\value{mv}}
\setcounter{mz}{\value{mxb}}
\setcounter{mv}{\value{myb}}
\addtocounter{mxb}{\value{mz}}
\addtocounter{myb}{\value{mv}}
\setcounter{mxc}{\value{mz}}
\setcounter{myc}{\value{mv}}
\addtocounter{mxb}{-\value{mv}}
\addtocounter{myb}{\value{mz}}
\addtocounter{mxc}{\value{mz}}
\addtocounter{myc}{\value{mv}}
\addtocounter{mx}{\value{cxa}}
\addtocounter{my}{\value{cya}}
\put(0,0){\begin{picture}(0,0)
\put(0,0) {\bezier{\value{fdots}}(0,0)(\value{mx},\value{my})(\value{mxa},\value{mya})} 
\put(\value{mxb},\value{myb}) {\bezier{\value{mdots}}(0,0)(\value{mv},-
\value{mz})(\value{myc},-\value{mxc})}
\end{picture}
}}

\newcommand{\curvedsolidline}[5]{
\setcounter{mdots}{#1}
\setcounter{mx}{#2}
\setcounter{my}{#3}
\setcounter{coun}{0}
\setcounter{mya}{0}
\setcounter{mxa}{0}
\setcounter{cya}{0}
\setcounter{cxa}{0}
\setcounter{fdots}{0}
\whiledo{#4 > \value{coun}}{\addtocounter{coun}{1} \addtocounter{fdots}{\value{mdots}}
\addtocounter{mxa}{\value{mx}} \addtocounter{mya}{\value{my}} }
\setcounter{coun}{0}
\whiledo{#5 > \value{coun}}{\addtocounter{coun}{1} \addtocounter{cxa}{-\value{cy}} \addtocounter{cya}{\value{cx}} }
\setcounter{cy}{\value{cya}}
\setcounter{cx}{\value{cxa}}
\addtocounter{cxa}{\value{cx}}
\addtocounter{cya}{\value{cy}}
\setcounter{my}{\value{mya}}
\setcounter{mx}{\value{mxa}}
\addtocounter{mxa}{\value{mx}}
\addtocounter{mya}{\value{my}}
\setcounter{my}{\value{mya}}
\setcounter{mx}{\value{mxa}}
\addtocounter{mxa}{\value{mx}}
\addtocounter{mya}{\value{my}}
\addtocounter{mx}{\value{cxa}}
\addtocounter{my}{\value{cya}}
\put(0,0){\begin{picture}(0,0)
\put(0,0) {\bezier{\value{fdots}}(0,0)(\value{mx},\value{my})(\value{mxa},\value{mya})} \end{picture}
}}

\newcounter{rx}
\newcounter{ry}

\newcounter{rxa}
\newcounter{rya}

\newcommand{\revmatterline}[6]{
\setcounter{rx}{#2}
\setcounter{ry}{#3}
\setcounter{rxa}{0}
\setcounter{rya}{0}
\setcounter{coun}{0}
\whiledo{#6 > \value{coun}}{\addtocounter{coun}{1} \addtocounter{rxa}{\value{rx}}
\addtocounter{rya}{\value{ry}} }
\addtocounter{rxa}{\value{rxa}}
\addtocounter{rya}{\value{rya}}
\addtocounter{rxa}{\value{rxa}}
\addtocounter{rya}{\value{rya}}
\put(0,0){\begin{picture}(0,0)
\put(\value{rxa},\value{rya}){\newmatterline{#1}{-#2}{-#3}{-#4}{-#5}{#6}}
\end{picture}}
}

\newcommand{\revcurvedmatterline}[7]{
\setcounter{rx}{#2}
\setcounter{ry}{#3}
\setcounter{rxa}{0}
\setcounter{rya}{0}
\setcounter{coun}{0}
\whiledo{#6 > \value{coun}}{\addtocounter{coun}{1} \addtocounter{rxa}{\value{rx}}
\addtocounter{rya}{\value{ry}} }
\addtocounter{rxa}{\value{rxa}}
\addtocounter{rya}{\value{rya}}
\addtocounter{rxa}{\value{rxa}}
\addtocounter{rya}{\value{rya}}
\put(0,0){\begin{picture}(0,0)
\put(\value{rxa},\value{rya}){\curvedmatterline{#1}{-#2}{-#3}{-#4}{-#5}{#6}{-#7}}
\end{picture}}
}

\newcounter{msxa}
\newcounter{msya}

\newcommand{\massprop}[6]{
\newmatterline{#1}{#2}{#3}{#4}{#5}{#6}
\setcounter{coun}{2}
\setcounter{msxa}{#6}
\addtocounter{msxa}{\value{msxa}}
\whiledo{\value{msxa} > \value{coun}}{\addtocounter{coun}{1} \addtocounter{mxb}{\value{mz} } \addtocounter{myb}{\value{mv}} }
\put(\value{mxb},\value{myb}) {\bezier{\value{mdots}}(0,0)(\value{mv},-
\value{mz})(\value{myc},-\value{mxc})}
}

\newcommand{\curvedmassprop}[7]{
\curvedmatterline{#1}{#2}{#3}{#4}{#5}{#6}{#7}
\setcounter{mz}{-#4}
\setcounter{mv}{-#5}
\setcounter{cz}{-\value{cz}}
\setcounter{cv}{-\value{cv}}
\setcounter{coun}{0}
\setcounter{cza}{0}
\setcounter{cva}{0}
\setcounter{myb}{0}
\setcounter{mxb}{0}
\whiledo{#6 > \value{coun}}{\addtocounter{coun}{1} \addtocounter{mxb}{\value{mz}} \addtocounter{myb}{\value{mv}} }
\setcounter{coun}{0}
\setcounter{disp}{-#7}
\ifthenelse{\value{disp} < 0}{ \setcounter{cv}{-\value{cv}} \setcounter{cz}{-\value{cz}} \setcounter{disp}{-\value{disp}} }{}
\whiledo{\value{disp} > \value{coun}}{\addtocounter{coun}{1} \addtocounter{cva}{\value{cz}} \addtocounter{cza}{-\value{cv}}}
\addtocounter{mxb}{\value{cza}}
\addtocounter{myb}{\value{cva}}
\setcounter{mz}{\value{mxb}}
\setcounter{mv}{\value{myb}}
\addtocounter{mxb}{\value{mz}}
\addtocounter{myb}{\value{mv}}
\setcounter{mz}{\value{mxb}}
\setcounter{mv}{\value{myb}}
\addtocounter{mxb}{\value{mz}}
\addtocounter{myb}{\value{mv}}
\setcounter{mxc}{\value{mz}}
\setcounter{myc}{\value{mv}}
\addtocounter{mxb}{-\value{mv}}
\addtocounter{myb}{\value{mz}}
\addtocounter{mxc}{\value{mz}}
\addtocounter{myc}{\value{mv}}
\setcounter{rx}{#2}
\setcounter{ry}{#3}
\setcounter{rxa}{0}
\setcounter{rya}{0}
\setcounter{coun}{0}
\whiledo{#6 > \value{coun}}{\addtocounter{coun}{1} \addtocounter{rxa}{\value{rx}}
\addtocounter{rya}{\value{ry}} }
\addtocounter{rxa}{\value{rxa}}
\addtocounter{rya}{\value{rya}}
\addtocounter{rxa}{\value{rxa}}
\addtocounter{rya}{\value{rya}}
\addtocounter{mxb}{\value{rxa}}
\addtocounter{myb}{\value{rya}}
\put(\value{mxb},\value{myb}) {\bezier{\value{mdots}}(0,0)(\value{mv},-
\value{mz})(\value{myc},-\value{mxc})}
}

\newcounter{sa}
\newcounter{sb}
\newcounter{sc}
\newcounter{sd}
\newcounter{se}
\newcounter{sf}
\newcounter{sg}

\newcommand{\solidline}[4]{
\setcounter{sa}{#1}
\setcounter{sb}{#2}
\setcounter{sc}{#3}
\setcounter{sd}{0}
\setcounter{se}{0}
\setcounter{coun}{0}
\setcounter{fdots}{0}
\whiledo{#4 > \value{coun}}{\addtocounter{coun}{1} \addtocounter{fdots}{\value{sa}}
\addtocounter{sd}{\value{sb}} \addtocounter{se}{\value{sc}} }
\setcounter{sb}{\value{sd}}
\setcounter{sc}{\value{se}}
\addtocounter{sd}{\value{sb}}
\addtocounter{se}{\value{sc}}
\setcounter{sb}{\value{sd}}
\setcounter{sc}{\value{se}}
\addtocounter{sd}{\value{sb}}
\addtocounter{se}{\value{sc}}
\put(0,0){\begin{picture}(0,0)
\ifthenelse{\value{sb} = 0}{ \ifthenelse{\value{sc} > 0} {\put(0,0) {\line(0,1){\value{se}}} }{
\put(0,0) {\line(0,-1){-\value{se}}}  }} { \ifthenelse{\value{sc} = 0} { \ifthenelse{\value{sb} >
0}{\put(0,0){\line(1,0){\value{sd}}} }{ \put(0,0){\line(-1,0){-\value{sd}}} } } {\put(0,0)
{\bezier{\value{fdots}}(0,0)(\value{sb},\value{sc})(\value{sd},\value{se})} } }
\end{picture}}
}

\newcommand{\gaugeline}[4]{
\setcounter{gx}{#1}
\setcounter{gy}{#2}
\setcounter{gxa}{\value{gx}}
\setcounter{gya}{\value{gy}}
\setcounter{gxb}{\value{gx}}
\setcounter{gyb}{\value{gy}}
\addtocounter{gxa}{-\value{gy}}
\addtocounter{gya}{\value{gx}}
\addtocounter{gxb}{\value{gx}}
\addtocounter{gyb}{\value{gy}}
\setcounter{gxc}{\value{gxb}}
\setcounter{gyc}{\value{gyb}}
\setcounter{gxd}{\value{gxa}}
\setcounter{gyd}{\value{gya}}
\addtocounter{gxd}{-\value{gyb}}
\addtocounter{gyd}{\value{gxb}}
\addtocounter{gxc}{\value{gxb}}
\addtocounter{gyc}{\value{gyb}}
\put(0,0){\begin{picture}(0,0)
\multiput(0,0)(\value{gxc}, \value{gyc}){#3}{%
\put(0,0){\begin{picture}(0,0)
\put(0,0){\bezier{#4}(0,0)(\value{gxa}, \value{gya})(\value{gxb}, \value{gyb})}
\put(0,0){\bezier{#4}(\value{gxb}, \value{gyb})(\value{gyd},-\value{gxd})%
(\value{gxc}, \value{gyc})}%
\end{picture}}
}
\end{picture}
}}

\newcounter{ga}
\newcounter{gb}
\newcounter{gxo}
\newcounter{gyo}
\newcounter{gecc}
\newcounter{gecctwo}

\newcounter{tempa}
\newcounter{tempb}
\newcounter{ma}
\newcounter{mb}

\newcommand{\mline}[2]{
\setcounter{cz}{#1}
\setcounter{cv}{#2}
\setcounter{tempa}{0}
\setcounter{tempb}{0}
\setcounter{coun}{0}
\whiledo{\value{stepup} > \value{coun}}{\addtocounter{coun}{1} \addtocounter{tempa}{\value{cz}} \addtocounter{tempb}{\value{cv}} }
\setcounter{cx}{\value{tempa}}
\setcounter{cy}{\value{tempb}}
\setcounter{ma}{0}
\setcounter{mb}{0}
\setcounter{coun}{0}
\whiledo{\value{stepup} > \value{coun}}{\addtocounter{coun}{1} \addtocounter{ma}{\value{tempa}} \addtocounter{mb}{\value{tempb}} }
\newmatterline{\value{dots}}{\value{ma}}{\value{mb}}{\value{tempa}}{\value{tempb}}{\value{its}}}

\newcommand{\cmline}[3]{
\setcounter{cz}{#1}
\setcounter{cv}{#2}
\setcounter{tempa}{0}
\setcounter{tempb}{0}
\setcounter{coun}{0}
\whiledo{\value{stepup} > \value{coun}}{\addtocounter{coun}{1} \addtocounter{tempa}{\value{cz}} \addtocounter{tempb}{\value{cv}} }
\setcounter{cx}{\value{tempa}}
\setcounter{cy}{\value{tempb}}
\setcounter{ma}{0}
\setcounter{mb}{0}
\setcounter{coun}{0}
\whiledo{\value{stepup} > \value{coun}}{\addtocounter{coun}{1} \addtocounter{ma}{\value{tempa}} \addtocounter{mb}{\value{tempb}} }
\curvedmatterline{\value{dots}}{\value{ma}}{\value{mb}}{\value{tempa}}{\value{tempb}}{\value{its}}{#3}}

\newcommand{\cbline}[3]{
\setcounter{cz}{#1}
\setcounter{cv}{#2}
\setcounter{tempa}{0}
\setcounter{tempb}{0}
\setcounter{coun}{0}
\whiledo{\value{stepup} > \value{coun}}{\addtocounter{coun}{1} \addtocounter{tempa}{\value{cz}} \addtocounter{tempb}{\value{cv}} }
\setcounter{cx}{\value{tempa}}
\setcounter{cy}{\value{tempb}}
\setcounter{ma}{0}
\setcounter{mb}{0}
\setcounter{coun}{0}
\whiledo{\value{stepup} > \value{coun}}{\addtocounter{coun}{1} \addtocounter{ma}{\value{tempa}} \addtocounter{mb}{\value{tempb}} }
\curvedmassprop{\value{dots}}{\value{ma}}{\value{mb}}{\value{tempa}}{\value{tempb}}{\value{its}}{#3}}

\newcommand{\crline}[3]{
\setcounter{cz}{#1}
\setcounter{cv}{#2}
\setcounter{tempa}{0}
\setcounter{tempb}{0}
\setcounter{coun}{0}
\whiledo{\value{stepup} > \value{coun}}{\addtocounter{coun}{1} \addtocounter{tempa}{\value{cz}} \addtocounter{tempb}{\value{cv}} }
\setcounter{cx}{\value{tempa}}
\setcounter{cy}{\value{tempb}}
\setcounter{ma}{0}
\setcounter{mb}{0}
\setcounter{coun}{0}
\whiledo{\value{stepup} > \value{coun}}{\addtocounter{coun}{1} \addtocounter{ma}{\value{tempa}} \addtocounter{mb}{\value{tempb}} }
\revcurvedmatterline{\value{dots}}{\value{ma}}{\value{mb}}{\value{tempa}}{\value{tempb}}{\value{its}}{-#3}}

\newcommand{\csline}[3]{
\setcounter{cx}{#1}
\setcounter{cy}{#2}
\setcounter{tempa}{0}
\setcounter{tempb}{0}
\setcounter{coun}{0}
\whiledo{\value{stepup} > \value{coun}}{\addtocounter{coun}{1} \addtocounter{tempa}{\value{cx}} \addtocounter{tempb}{\value{cy}} }
\setcounter{cx}{\value{tempa}}
\setcounter{cy}{\value{tempb}}
\setcounter{ma}{0}
\setcounter{mb}{0}
\setcounter{coun}{0}
\whiledo{\value{stepup} > \value{coun}}{\addtocounter{coun}{1} \addtocounter{ma}{\value{tempa}} \addtocounter{mb}{\value{tempb}} }
\curvedsolidline{\value{dots}}{\value{ma}}{\value{mb}}{\value{its}}{#3}}

\newcommand{\rline}[2]{
\setcounter{cx}{#1}
\setcounter{cy}{#2}
\setcounter{tempa}{0}
\setcounter{tempb}{0}
\setcounter{coun}{0}
\whiledo{\value{stepup} > \value{coun}}{\addtocounter{coun}{1} \addtocounter{tempa}{\value{cx}} \addtocounter{tempb}{\value{cy}} }
\setcounter{cx}{\value{tempa}}
\setcounter{cy}{\value{tempb}}
\setcounter{ma}{0}
\setcounter{mb}{0}
\setcounter{coun}{0}
\whiledo{\value{stepup} > \value{coun}}{\addtocounter{coun}{1} \addtocounter{ma}{\value{tempa}} \addtocounter{mb}{\value{tempb}} }
\revmatterline{\value{dots}}{\value{ma}}{\value{mb}}{\value{tempa}}{\value{tempb}}{\value{its}}}

\newcommand{\gline}[2]{\setcounter{cx}{#1}
\setcounter{cy}{#2}
\setcounter{tempa}{0}
\setcounter{tempb}{0}
\setcounter{coun}{0}
\whiledo{\value{stepup} > \value{coun}}{\addtocounter{coun}{1} \addtocounter{tempa}{\value{cx}} \addtocounter{tempb}{\value{cy}} }
\setcounter{cx}{\value{tempa}}
\setcounter{cy}{\value{tempb}}
\setcounter{ma}{0}
\setcounter{mb}{0}
\setcounter{coun}{0}
\whiledo{\value{stepup} > \value{coun}}{\addtocounter{coun}{1} \addtocounter{ma}{\value{tempa}} \addtocounter{mb}{\value{tempb}} }
\gaugeline{\value{ma}}{\value{mb}}{\value{its}}{\value{gdots}}}

\newcommand{\bline}[2]{
\setcounter{cx}{#1}
\setcounter{cy}{#2}
\setcounter{tempa}{0}
\setcounter{tempb}{0}
\setcounter{coun}{0}
\whiledo{\value{stepup} > \value{coun}}{\addtocounter{coun}{1} \addtocounter{tempa}{\value{cx}} \addtocounter{tempb}{\value{cy}} }
\setcounter{cx}{\value{tempa}}
\setcounter{cy}{\value{tempb}}
\setcounter{ma}{0}
\setcounter{mb}{0}
\setcounter{coun}{0}
\whiledo{\value{stepup} > \value{coun}}{\addtocounter{coun}{1} \addtocounter{ma}{\value{tempa}} \addtocounter{mb}{\value{tempb}} }
\massprop{\value{dots}}{\value{ma}}{\value{mb}}{\value{tempa}}{\value{tempb}}{\value{its}}}

\newcommand{\sline}[2]{
\setcounter{cx}{#1}
\setcounter{cy}{#2}
\setcounter{tempa}{0}
\setcounter{tempb}{0}
\setcounter{coun}{0}
\whiledo{\value{stepup} > \value{coun}}{\addtocounter{coun}{1} \addtocounter{tempa}{\value{cx}} \addtocounter{tempb}{\value{cy}} }
\setcounter{cx}{\value{tempa}}
\setcounter{cy}{\value{tempb}}
\setcounter{ma}{0}
\setcounter{mb}{0}
\setcounter{coun}{0}
\whiledo{\value{stepup} > \value{coun}}{\addtocounter{coun}{1} \addtocounter{ma}{\value{tempa}} \addtocounter{mb}{\value{tempb}} }
\solidline{\value{dots}}{\value{ma}}{\value{mb}}{\value{its}}}

\newcommand{\circularline}[5]{
\setcounter{sb}{#1}
\setcounter{sc}{#2}
\setcounter{sd}{0}
\setcounter{se}{0}
\setcounter{coun}{0}
\whiledo{#3 > \value{coun}}{\addtocounter{coun}{1}
\addtocounter{sd}{\value{sb}} \addtocounter{se}{\value{sc}} }
\setcounter{sb}{\value{sd}}
\setcounter{sc}{\value{se}}
\addtocounter{sd}{\value{sb}}
\addtocounter{se}{\value{sc}}
\setcounter{sb}{\value{sd}}
\setcounter{sc}{\value{se}}
\addtocounter{sd}{\value{sb}}
\addtocounter{se}{\value{sc}}
\put(0,0){\begin{picture}(\value{sb},\value{sc})
\put(\value{sb},\value{sc}){\circle{#4}\makebox(0,0){#5}}
\end{picture}}
}

\newcommand{\ccline}[4]{
\setcounter{cx}{#1}
\setcounter{cy}{#2}
\setcounter{tempa}{0}
\setcounter{tempb}{0}
\setcounter{coun}{0}
\whiledo{\value{stepup} > \value{coun}}{\addtocounter{coun}{1} \addtocounter{tempa}{\value{cx}} \addtocounter{tempb}{\value{cy}} }
\setcounter{cx}{\value{tempa}}
\setcounter{cy}{\value{tempb}}
\setcounter{ma}{0}
\setcounter{mb}{0}
\setcounter{coun}{0}
\whiledo{\value{stepup} > \value{coun}}{\addtocounter{coun}{1} \addtocounter{ma}{\value{tempa}} \addtocounter{mb}{\value{tempb}} }
\circularline{\value{ma}}{\value{mb}}{\value{its}}{#3}{#4}}

%% file: spldefs.tex
\newcommand{\piccy}[2]{\begin{picture}(185500,#2)
\put(0,0){\framebox(184000,#2){The picture}}
\end{picture}}
\newcommand{\sect}[1]{Section~\ref{sec:#1}}
\newcommand{\app}[1]{\ref{sec:#1}}
\newcommand{\fig}[1]{Figure~\ref{fig:#1}}
\newcommand{\tab}[1]{Table~\ref{tab:#1}}
\newcommand{\eqnnum}[1]{(\ref{eqn:#1})}
\newcommand{\eqn}[1]{Eq.~\eqnnum{#1}}
\newcommand{\bibnum}[1]{\cite{bib:#1}}
\newcommand{\bib}[1]{Ref. \bibnum{#1}}
\newcommand{\beq}{\begin{eqnarray}}
\newcommand{\eeq}{\end{eqnarray}}
\newcommand{\bed}{\begin{displaymath}}
\newcommand{\eed}{\end{displaymath}}

\newcommand{\wzm}{Wess-Zumino model}
\newcommand{\abe}{Abelian}
\newcommand{\sumfac}[1]{\eta(#1)}

\newenvironment{note}[1]{\textbf{#1}}{}
\newcommand{\bnt}{\begin{note}}
\newcommand{\ent}{\end{note}}

\newcommand{\ksqu}{\ensuremath{r}}
\newcommand{\spct}{\ensuremath{x}}
\newcommand{\veee}{\ensuremath{V}}
\newcommand{\kay}{\ensuremath{K}}
\newcommand{\eff}[1]{\ensuremath{F_{#1}}}
\newcommand{\kaysb}[1]{\ensuremath{K^{#1}}}
\newcommand{\effsb}[2]{\ensuremath{F_{#1}^{#2}}}
\newcommand{\weee}{\ensuremath{W_{A}}}
\newcommand{\wsqu}{\ensuremath{W^{2}}}
\newcommand{\wsquind}{\ensuremath{W^{A} W_{A}}}
\newcommand{\tht}[1]{\ensuremath{\theta_{#1}}}

\newcommand{\cov}[1]{\ensuremath{D_{#1}}}
\newcommand{\upcov}[1]{\ensuremath{D^{#1}}}
\newcommand{\covsqu}{\ensuremath{D^{2}}}
\newcommand{\covbar}[1]{\ensuremath{\bar{D}_{\dot{#1}}}}
\newcommand{\upcovbar}[1]{\ensuremath{\bar{D}^{\dot{#1}}}}
\newcommand{\covbarsqu}{\ensuremath{\bar{D}^{2}}}
\newcommand{\fye}{\ensuremath{\Phi}}
\newcommand{\fyebar}{\ensuremath{\bar{\Phi}}}
\newcommand{\dfye}{\ensuremath{\cov{A} \Phi}}
\newcommand{\dfyebar}{\ensuremath{\covbar{B} \bar{\Phi}}}
\newcommand{\kayl}{K\"{a}hlerian}
\newcommand{\fyebarfye}[1]{\ensuremath{\bar{\Phi}_{#1} \Phi_{#1}}}
\newcommand{\dalg}{$D$-algebra}
\newcommand{\supth}[1]{$#1^{\mathit{th}}$}
\newcommand{\psye}{\ensuremath{\Psi}}
\newcommand{\psyebar}{\ensuremath{\bar{\Psi}}}
\newcommand{\dfs}[2]{\, d^{4}#1_{#2} \, }
\newcommand{\dz}[1]{\, d^{#1}z \, }
\newcommand{\df}[1]{\, d^{4}#1\, }
\newcommand{\dt}[1]{d^{2}#1\, }
\newcommand{\twoverts}[3]{\ensuremath{\left\{ \lambda \Phi_{#1} \bar{D}^{2}_{#1} D_{#2}^{2} \delta_{#1 #2} \right\} \lambda \bar{\Phi}_{#2} \delta_{#2 #3}}}
\newcommand{\sectwoverts}[2]{ \ensuremath{\left\{ \bar{\Phi}_{#1} \Phi_{#1} \bar{D}^{2}_{#1} D_{#2}^{2} \delta_{#1 #2} \right\} }}
\newcommand{\nobsectwoverts}[2]{ \ensuremath{ \bar{\Phi}_{#1} \Phi_{#1} \bar{D}^{2}_{#1} D_{#2}^{2} \delta_{#1 #2} }}
\newcommand{\thdtwoverts}[2]{ \ensuremath{\bar{\Phi}_{#1} \Phi_{#1} \bar{D}^{2}_{#2} D_{#2}^{2}}}
\newcommand{\genvert}[1]{ \ensuremath{#1 \; D^{2} \bar{D}^{2}}}
\newcommand{\blockphi}{ \ensuremath{\upcov{A} \Phi \cov{A} \Phi \upcovbar{B} \bar{\Phi} \covbar{B} \bar{\Phi} }}
\newcommand{\blockpsye}{ \ensuremath{\upcov{A} \Psi \cov{A} \Psi \upcovbar{B} \bar{\Psi} \covbar{B} \bar{\Psi} }}
\newcommand{\shortblockphi}{ \ensuremath{(D \Phi \bar{D} \bar{\Phi})^{2} }}
\newcommand{\dtwogamma}[1]{\ensuremath{\frac{\partial^{2} #1}{\partial \fye \partial \fyebar} } }
\newcommand{\dtwogammaind}[3]{\ensuremath{\frac{\partial^{2} #1}{\partial \fyebar^{#2} \partial \fye_{#3}} } }

\newcommand{\biggam}{\ensuremath{\Gamma}}
\newcommand{\biggamzero}{\ensuremath{\Gamma(0)}}
\newcommand{\biggamsp}[1]{\ensuremath{\Gamma^{(#1)}}}
\newcommand{\biggamspsb}[2]{\ensuremath{\Gamma^{(#1)}_{#2}}}
\newcommand{\biggamsb}[1]{\ensuremath{\Gamma_{#1}}}
\newcommand{\biggammomordsp}[2]{\ensuremath{\Gamma^{(#1)}(p_{i})[#2]}}
\newcommand{\biggammom}{\ensuremath{\Gamma(p_{i})}}
\newcommand{\biggamord}[1]{\ensuremath{\Gamma[#1]}}
\newcommand{\biggamordsp}[2]{\ensuremath{\Gamma^{(#1)}[#2]}}
\newcommand{\biggamordspsb}[3]{\ensuremath{\Gamma^{(#1)}_{#3}[#2]}}
\newcommand{\biggamordsb}[2]{\ensuremath{\Gamma^{(#1)}_{#2}}}

\newcommand{\renlims}[1]{\ensuremath{ \left. #1 \right|_{ \stackrel{ \scriptstyle \fye  =  \fye_{0}}{  \fyebar  =  \fyebar_{0}}  }  } }

\newcommand{\dtwogammapsye}[1]{\ensuremath{\frac{\partial^{2} #1}{\partial \psye \partial \psyebar} } }
\newcommand{\renlimspsye}{\ensuremath{ \right|_{ \stackrel{ \scriptstyle \psye  =  \psye_{0}}{  \psyebar  =  \psyebar_{0}}  }  } }

\newcommand{\ffprop}{\ensuremath{\langle \fye \fye \rangle}}
\newcommand{\bbprop}{\ensuremath{\langle \fyebar \fyebar \rangle}}
\newcommand{\vvprop}{\ensuremath{\langle V V \rangle}}
\newcommand{\bfprop}{\ensuremath{\langle \fyebar \fye \rangle}}
\newcommand{\fbprop}{\ensuremath{\langle \fye \fyebar \rangle}}
\newcommand{\ymt}{N = 1 U(1) gauge theory}
\newcommand{\none}{\ensuremath{N = 1}}
\newcommand{\ntwo}{\ensuremath{N = 2}}
\newcommand{\nabymt}{Yang-Mills theory}
\newcommand{\uone}{U(1)}
\newcommand{\figpartnum}[1]{\ref{itm:#1}}
\newcommand{\figfullnum}[2]{\ref{fig:#1}\figpartnum{#2}}
\newcommand{\ymvrt}[1]{Figure~\ref{fig:ymvertopts}(#1)}
\newcommand{\figfull}[2]{Figure~\figfullnum{#1}{#2}}
\newcommand{\delfns}[2]{\ensuremath{\delta_{#1 #2}}-function}
\newcommand{\delfn}{\ensuremath{\delta}-function}
\newcommand{\bep}{\begin{picture}}
\newcommand{\eep}{\end{picture}}
\renewcommand{\theenumi}{(\alph{enumi})}
\renewcommand{\labelenumi}{\theenumi}
\newcommand{\bef}{\begin{figure}}
\newcommand{\eef}{\end{figure}}
\newcommand{\bet}{\begin{table}}
\newcommand{\eet}{\end{table}}
\newcommand{\kap}[1]{\caption{#1}}
\newcommand{\flab}[1]{\label{fig:#1}}
\newcommand{\tlab}[1]{\label{tab:#1}}
\newcommand{\elab}[1]{\label{eqn:#1}}
\newcommand{\ilab}[1]{\label{itm:#1}}
\newcommand{\slab}[1]{\label{sec:#1}}

\newcommand{\extgen}{\ensuremath{G}}
\newcommand{\dub}{\covsqu \covbarsqu}
\newcommand{\revdub}{\covbarsqu \covsqu}

\newcommand{\fulltee}[3]{{{{(T^{#1})}_{#2}}^{#3}}}
\newcommand{\mat}[3]{{#1}^{#2 #3}}
\newcommand{\tee}[1]{T^{#1}}
\newcommand{\trce}{\mathrm{Tr}}
\newcommand{\trcearg}[1]{\mathrm{Tr} \left( #1 \right)}

\newcommand{\why}{\ensuremath{\lambda^{2} \fyebar \fye}}
\newcommand{\masswhy}{\ensuremath{\frac{\lambda^{2} \fyebar \fye x}{y^{2}}}}

\newcommand{\hlf}{\ensuremath{\frac{1}{2} }}
\newcommand{\intone}{\ensuremath{\int \frac{dx \, \df{\theta} }{16 \pi^{2}} \, }}
\newcommand{\inttwo}{\ensuremath{\int \frac{dx \, \df{\theta} }{(16 \pi)^{2}} \, }}
\newcommand{\bfone}{\ensuremath{\beta \fyebar \fye}}
\newcommand{\bfoneb}{\ensuremath{\left( \beta \fyebar \fye \right) }}
\newcommand{\bftwo}{\ensuremath{\left( 1 - \beta \fyebar \fye \right)}}
\newcommand{\gvone}{\ensuremath{gV}}
\newcommand{\gvtwo}{\ensuremath{g^{2} V^{2}}}
\newcommand{\vdubv}{\ensuremath{V} \dub \ensuremath{( V )}}

\newcommand{\splteqn}[1]{ $ \begin{array}{@{}l@{}} \displaystyle 
 #1 
\end{array} $ }
\newcommand{\nwln}{ \\[10 pt] \displaystyle } 

\newcommand{\matone}[2]{\ensuremath{ \beta \fyebar \tee{#1} V \tee{#2} \fye }}
\newcommand{\mattwo}[2]{\ensuremath{ \frac{\beta}{2} \fyebar \left\{ V,\tee{#1} \right\} \tee{#2} \fye }}
\newcommand{\matthr}[2]{\ensuremath{ \frac{\beta}{2} \fyebar \tee{#1} \left\{ V, \tee{#2} \right\}  \fye }}
\newcommand{\matfou}[2]{\ensuremath{ \frac{\beta}{3!} \fyebar \left\{ V, \tee{#1} , \tee{#2} \right\}  \fye }}
\newcommand{\dubmatone}[2]{\ensuremath{ \beta \fyebar \tee{#1} \dub(V) \tee{#2} \fye }}
\newcommand{\revdubmatone}[2]{\ensuremath{ \beta \fyebar \tee{#1} \revdub(V) \tee{#2} \fye }}
\newcommand{\dubmattwo}[2]{\ensuremath{ \frac{\beta}{2} \fyebar \left\{ \dub(V),\tee{#1} \right\} \tee{#2} \fye }}
\newcommand{\revdubmattwo}[2]{\ensuremath{ \frac{\beta}{2} \fyebar \left\{ \revdub(V),\tee{#1} \right\} \tee{#2} \fye }}
\newcommand{\dubmatthr}[2]{\ensuremath{ \frac{\beta}{2} \fyebar \tee{#1} \left\{ \dub(V), \tee{#2} \right\}  \fye }}
\newcommand{\dubmatfou}[2]{\ensuremath{ \frac{\beta}{3!} \fyebar \left\{ \dub(V), \tee{#1} , \tee{#2} \right\}  \fye }}
\newcommand{\revdubmatthr}[2]{\ensuremath{ \frac{\beta}{2} \fyebar \tee{#1} \left\{ \revdub(V), \tee{#2} \right\}  \fye }}
\newcommand{\revdubmatfou}[2]{\ensuremath{ \frac{\beta}{3!} \fyebar \left\{ \revdub(V), \tee{#1} , \tee{#2} \right\}  \fye }}
\newcommand{\matfiv}[2]{\ensuremath{ \beta \fyebar \tee{#1} V^{2} \tee{#2} \fye }}
\newcommand{\matsix}[2]{\ensuremath{ \frac{\beta}{3!} \fyebar \left\{ V, V, \tee{#1} \right\} \tee{#2}  \fye }}
\newcommand{\matsev}[2]{\ensuremath{ \frac{\beta}{3!} \fyebar \tee{#1} \left\{ V, V, \tee{#2} \right\}  \fye }}
\newcommand{\mateig}[2]{\ensuremath{ \frac{\beta}{4!} \fyebar \left\{ V, V, \tee{#1} , \tee{#2} \right\}  \fye }}
\newcommand{\matnin}[2]{\ensuremath{ \frac{\beta}{2} \fyebar \left\{ V,\tee{#1} \right\} V \tee{#2} \fye }}
\newcommand{\matten}[2]{\ensuremath{ \frac{\beta}{2} \fyebar \tee{#1} V \left\{ V, \tee{#2} \right\}  \fye }}
\newcommand{\matele}[2]{\ensuremath{ \frac{\beta}{4} \fyebar \left\{ \tee{#1}, V \right\} \left\{ V, \tee{#2} \right\}  \fye }}
\newcommand{\revdubmatele}[2]{\ensuremath{ \frac{\beta}{4} \fyebar \left\{ \tee{#1}, \revdub(V) \right\} \left\{ V, \tee{#2} \right\}  \fye }}
\newcommand{\revdubmatnin}[2]{\ensuremath{ \frac{\beta}{2} \fyebar \left\{ V,\tee{#1} \right\} \revdub(V) \tee{#2} \fye }}
\newcommand{\revdubmatten}[2]{\ensuremath{ \frac{\beta}{2} \fyebar \tee{#1} \revdub(V) \left\{ V, \tee{#2} \right\}  \fye }}
\newcommand{\dubmatnin}[2]{\ensuremath{ \frac{\beta}{2} \fyebar \left\{ \dub(V),\tee{#1} \right\} V \tee{#2} \fye }}
\newcommand{\dubmatten}[2]{\ensuremath{ \frac{\beta}{2} \fyebar \tee{#1} V \left\{ \dub(V), \tee{#2} \right\}  \fye }}
\newcommand{\emme}[2]{\ensuremath{ \mat{M}{#1}{#2} }}
\newcommand{\enne}[2]{\ensuremath{ \mat{N}{#1}{#2} }}
\newcommand{\bee}[2]{\ensuremath{ \mat{B}{#1}{#2} }}
\newcommand{\cee}[2]{\ensuremath{ \mat{\left( \frac{1}{1-A} \right) }{#1}{#2} }}
\newcommand{\cbm}[2]{\ensuremath{ \mat{\left( \frac{1}{1-A} BM \right) }{#1}{#2} }}

\newcommand{\ord}[1]{\ensuremath{ O \left( #1 \right)  }}

\remark{
\makeatletter 
         \renewcommand\theequation{\thesection.\arabic{equation}}
         \@addtoreset{equation}{section}
\makeatother
}

\newcommand{\myappendix}{
\appendix
\addtocontents{toc}{\protect\newsec}}

\newcommand{\appsection}[1]{
\renewcommand{\thesection}{Appendix~\Alph{section}}
\section{#1}
\renewcommand{\thesection}{\Alph{section}}}

\newlength{\appwidth}
\settowidth{\appwidth}{Appendix A }

\makeatletter
\newcommand{\newsec}{
\renewcommand{\l@section}[2]{%
    \addpenalty{\@secpenalty}%
    \addvspace{1.0em \@plus\p@}%
    \setlength\@tempdima{1.5em}%
    \addtolength{\@tempdima}{\appwidth}%
    \begingroup
    \parindent \z@ \rightskip \@pnumwidth
    \parfillskip -\@pnumwidth
    \leavevmode \bfseries
    \advance\leftskip\@tempdima
    \hskip -\leftskip
    ##1\nobreak\hfil \nobreak\hbox to\@pnumwidth{\hss ##2}\par
   \endgroup}
}
\makeatother

%% file: splpc01.tex
\begin{picture}(122400,122400)(-61200,-61200)
\setcounter{dots}{10}
\put(-23562,-17119){\cmline{-8}{24}{10}}
\put(-23562,17441){\cmline{20}{15}{10}}
\put(8838,-27919){\cmline{-25}{0}{10}}
\put(8838,28241){\cmline{20}{-15}{10}}
\put(-9162,-27919){\crline{-20}{15}{10}}
\put(-29322,161){\crline{8}{24}{10}}
\put(-9162,28241){\crline{25}{0}{10}}
\put(26118,-8479){\circle*{1000}}
\put(26118,8801){\circle*{1000}}
\put(16038,-22519){\circle*{1000}}
\put(-9162,-27919){\sline{-8}{-24}}
\put(-23562,-17119){\mline{-20}{-15}}
\put(-29322,161){\sline{-25}{0}}
\put(-23562,17441){\mline{-20}{15}}
\put(-9162,28241){\sline{-8}{24}}
\put(8838,28241){\mline{8}{24}}
\put(-33177,10780){\makebox(0,0){$k$}}
\put(-56125,0){\makebox(0,0){$p_{1}=0$}}
\put(-45406,32989){\makebox(0,0){$p_{2}=0$}}
\put(-17343,53378){\makebox(0,0){$p_{3}=0$}}
\put(17343,53378){\makebox(0,0){$p_{4}=0$}}
\put(-17343,-53378){\makebox(0,0){$p_{2n-1}=0$}}
\put(-45406,-32989){\makebox(0,0){$p_{2n}=0$}}
\end{picture}

%% file: spltb01.tex
\[
\remark{\displaystyle}
\begin{array}{@{}|r|r|l|}
\hline
i & c_{i} &\multicolumn{1}{|c|}{  I_{i} } \\
\hline
               1 & 1 &  \remark{\displaystyle} \sum_{n=2}^{\infty} \sumfac{n}    
                  \sum_{p=1}^{n-1}    
                   {G^{-p + n}}\,D^{2}\bar{D}^{2}({G^p})    \\
              2 & 1 &  \remark{\displaystyle} \sum_{n=3}^{\infty}  \sumfac{n}    
                 \sum_{p=1}^{n-2 }    
                  {G^{-1 - p + n}}\,\bar{D}^{2}(G\,D^{2}({G^p})) 
                     \\
             3 & -2 &  \remark{\displaystyle} \sum_{n=3}^{\infty} \sumfac{n}    
                \sum_{p=1}^{n-2}    
                 {G^{n-p-1}}\,
                  \bar{D}^{\dot{B}}(G\,D^{2}\bar{D}_{\dot{B}}({G^p}))  
                   \\
            4 & -2 &  \remark{\displaystyle} \sum_{n=3}^{\infty} \sumfac{n}    
               \sum_{p=1}^{n-2}    
                {G^{-1 - p + n}}\,D^{A}\bar{D}^{2}(G\,D_{A}({G^p}))
                       \\
            5 & -2 &  \remark{\displaystyle} 
             \sum_{n=4}^{\infty} \sumfac{n}    
              \sum_{p=1}^{n-3}    
               {G^{-2 - p + n}}\,
                \bar{D}^{\dot{B}}(G\,D^{A}\bar{D}_{\dot{B}}(G\,D_{A}({G^p})))
                      \\
           6 & -2 &  
            \remark{\displaystyle} \sum_{n=5}^{\infty} \sumfac{n}    
             \sum_{q=3}^{n-2} \sum_{p=1}^{q-2}    
              {G^{n-q-1}}\,
               \bar{D}^{\dot{B}}(
                G\,D^{A}({G^{q-p-1}}\,\bar{D}_{\dot{B}}(G\,D_{A}({G^p})))
                )      \\
          7 & -2 &  
           \remark{\displaystyle} \sum_{n=4}^{\infty} \sumfac{n}    
            \sum_{q=3}^{-1 + n} \sum_{p=1}^{q-2}    
             {G^{-q + n}}\,
              D^{A}\bar{D}^{\dot{B}}(
               {G^{-1 - p + q}}\,\bar{D}_{\dot{B}}(G\,D_{A}({G^p})))  
               \\
        8 & -2 &  \remark{\displaystyle} \sum_{n=4}^{\infty} \sumfac{n}    
           \sum_{q=2}^{n-2} \sum_{p=1}^{q-1}    
            {G^{-1 - q + n}}\,
             \bar{D}^{\dot{B}}(G\,
               D^{A}({G^{-p + q}}\,D_{A}\bar{D}_{\dot{B}}({G^p})))  
              \\
       9 & -2 &  \remark{\displaystyle} \sum_{n=3}^{\infty} \sumfac{n}    
          \sum_{q=2}^{-1 + n} \sum_{p=1}^{q-1}    
           {G^{-q + n}}\,D^{A}\bar{D}^{\dot{B}}(
             {G^{-p + q}}\,D_{A}\bar{D}_{\dot{B}}({G^p}))      \\
\hline
\end{array}
\]

%% file: splpc99.tex
\begin{picture}(50400,50400)(-25200,-25200)
\setcounter{dots}{10}
\put(-9000,-9000){\cmline{0}{25}{25}}
\put(9000,9000){\crline{0}{-25}{25}}
\put(-9000,9000){\begin{picture}(0,0)
\put(0,0){\bezier{\value{gdots}}(0,0)(111,1408)(1519,1297)
\bezier{\value{gdots}}(1519,1297)(2892,967)(3222,2341)
\bezier{\value{gdots}}(3222,2341)(3762,3645)(5067,3105)
\bezier{\value{gdots}}(5067,3105)(6271,2367)(7009,3571)
\bezier{\value{gdots}}(7009,3571)(7926,4645)(9000,3728)
\bezier{\value{gdots}}(9000,3728)(9917,2654)(10991,3571)
\bezier{\value{gdots}}(10991,3571)(12195,4309)(12933,3105)
\bezier{\value{gdots}}(12933,3105)(13474,1800)(14778,2341)
\bezier{\value{gdots}}(14778,2341)(16152,2670)(16481,1297)
\bezier{\value{gdots}}(16481,1297)(16592,-111)(18000,0)
}\end{picture}
}
\put(9000,-9000){\begin{picture}(0,0)
\put(0,0){\bezier{\value{gdots}}(0,0)(-111,-1408)(-1519,-1297)
\bezier{\value{gdots}}(-1519,-1297)(-2892,-967)(-3222,-2341)
\bezier{\value{gdots}}(-3222,-2341)(-3762,-3645)(-5067,-3105)
\bezier{\value{gdots}}(-5067,-3105)(-6271,-2367)(-7009,-3571)
\bezier{\value{gdots}}(-7009,-3571)(-7926,-4645)(-9000,-3728)
\bezier{\value{gdots}}(-9000,-3728)(-9917,-2654)(-10991,-3571)
\bezier{\value{gdots}}(-10991,-3571)(-12195,-4309)(-12933,-3105)
\bezier{\value{gdots}}(-12933,-3105)(-13474,-1800)(-14778,-2341)
\bezier{\value{gdots}}(-14778,-2341)(-16152,-2670)(-16481,-1297)
\bezier{\value{gdots}}(-16481,-1297)(-16592,111)(-18000,0)
}\end{picture}
}
\put(-9000,9000){\mline{-18}{18}}
\put(9000,9000){\mline{18}{18}}
\put(9000,-9000){\sline{18}{-18}}
\put(-9000,-9000){\sline{-18}{-18}}
\end{picture}

%% file: spl.bbl
\begin{thebibliography}{99}
\bibitem{bib:colewein} S. Coleman and E. Weinberg, Phys Rev. D  (1973) , vol. 7 no. 6,  pp. 1888-
-1910.
\bibitem{bib:buchbinder} I. Buchbinder, S. Kuzenko and J. Yarevskaya, Nucl. Phys {\bf B411}
(1994) pp. 665--692.
\bibitem{bib:westsusy} P. West, {\em Introduction to Supersymmetry and Supergravity } (World
Scientific, 1990 Extended Second Edition).
\remark{\bibitem{bib:sohnius} M. Sohnius, Phys Rep.  {\bf 128} (1985) , pp. 39- 204.}
\remark{\bibitem{bib:diagramma} G. 't Hooft and M. Veltman {\em Diagrammar } (CERN 73-9, Lab
1, Theoretical Studies Division, GENEVA 1973).}
\bibitem{bib:superspace} S. J. Gates, M.T. Grisaru, M. Ro\u{c}ek and W. Siegel, {\em
Superspace, or 1001 lessons in supersymmetry. }, Frontiers in Physics 58 (Addison Wesley,
1983).
\remark{\bibitem{bib:math} S. Wolfram, {\em Mathematica}, \copyright\  Wolfram Research Inc.}
\bibitem{bib:seib} S. J. Gates, Jr., Nucl. Phys. B238 (1984) 349, G. Sierra and P.K. Townsend, {\em Supersymmetry and Supergravity} (1983), Ed. B. Milewski , p. 396.
\bibitem{bib:wgr} B. deWit, M.T. Grisaru, M. Ro\u{c}ek, hep-th/9601115
\bibitem{bib:gru} M.T. Grisaru, M. Ro\u{c}ek, R. Unge hep-th/9605149
\end{thebibliography}
